\documentclass[useAMS,usenatbib]{mn2e}
\input psfig.sty

\def\msun{\thinspace\hbox{$\hbox{M}_{\odot}$}}

\title[An old quasar in a young dark energy-dominated universe?]
{An old quasar in a young dark energy-dominated universe?}
\author[Fria\c{c}a, Alcaniz, and Lima]{A. C. S.
Fria\c{c}a$^{1}$\thanks{E-mail: amancio@astro.iag.usp.br} J. S.
Alcaniz$^{2}$\thanks{E-mail: alcaniz@on.br} J. A. S.
Lima$^{1,3}$\thanks{E-mail: limajas@astro.iag.usp.br}\\
$^{1}$Instituto de Astronomia, Geof\'{\i}sica e Ci\^encias
Espaciais, Universidade de S\~ao Paulo, 05508-900 S\~ao Paulo -
SP, Brazil\\ $^{2}$Observat\'orio Nacional, Rua General Jos\'e Cristino 77, 20921-400, S\~ao
Cristov\~ao, Rio de Janeiro - RJ, Brazil\\ $^{3}$Departamento de F\'{\i}sica, Universidade
Federal do Rio Grande do Norte, C.P. 1641, 59072-970 Natal - RN,
Brazil}

\begin{document}

\date{Accepted ; Received}

\pagerange{\pageref{firstpage}--\pageref{lastpage}} \pubyear{2004}

\maketitle

\label{firstpage}

\begin{abstract}

Dark energy is the invisible fuel that seems to drive the current
acceleration of the Universe. Its presence, which is inferred from
an impressive convergence of high-quality observational results
along with some apparently sucessful theoretical predictions, is
also supported by the current estimates of the age of the Universe
from dating of local and high-$z$ objects. In this paper we test
the viability of several dark energy scenarios in the light of the
age estimates of the high redshift ($z=3.91$) quasar APM 08279+5255.
Using a chemodinamical model for the evolution of spheroids, 
we first reevaluate its current estimated age, as
given by Hasinger et al. (2002). 
An age of 2.1 Gyr is set by the condition that 
Fe/O abundance ratio (normalized to solar values) of the model reaches 3.3,
which is the best fit value obtained in the above reference.
In the detailed chemodynamical modelling, the iron enrichment
defines three relevant time scales: (i) $\sim 0.3$ Gyr for the
central region of the galaxy housing the quasar to reach a solar iron abundance;
(ii) $\sim 1$ Gyr for the Fe/O abundance ratio to reach the solar value;
(iii) $\sim 2$ Gyr for a highly suprasolar Fe/O abundance ratio
(Fe/O=2.5, suggested by the quasar APM 08279+5255).
Therefore, a high value of the Fe/O abundance ratio for a quasar
is a strong evidence that the quasar is old,
which represents a severe constraint for cosmological scenarios. 
It is shown that for the currently accepted value of
the matter density parameter,
most of the existing dark energy
scenarios cannot accomodate this old high redshift object unless
the Hubble parameter is as low as $H_o = 58$
$\rm{km.s^{-1}.Mpc^{-1}}$, as recently advocated by Sandage and collaborators.
Even considering less stringent age limits,
only cosmological models that predicts a considerably old Universe at high-$z$
can be compatible with the existence of this object.
This is the case of the conventional $\Lambda$CDM scenario
and some specific classes of brane world cosmologies.

\end{abstract}


\section{Introduction}

The idea of a dark energy-dominated universe is a direct
consequence of an impressive convergence of independent
observational results. Along with distance measurements from
high-redshift type Ia supernovae (SNe Ia), CMB anisotropies and
clustering estimates, one of the most presssing piece of data
which motivated such an idea involves the estimates of the age of
the Universe ($t_{\rm{U}}$). Since $t_{\rm{U}}$ at any stage of
the evolution of the Universe must necessarily be greater than the
age of any object within it, dark energy helps explain the current
dating of globular clusters, which indicates $t_{\rm{U}} \geq 13$
Gyr, by allowing a period of cosmic acceleration and leading to a
larger expansion age. For the widely accepted current value of the
Hubble parameter, i.e., $H_o = 72 \pm 8$ $\rm{km.s^{-1}.Mpc^{-1}}$
(Freedman et al. 2001), no flat CDM model without dark energy
(whose age prediction is $t_{\rm{U}} = {2 \over 3}H_o^{-1}$) may
be compatible either with the direct age estimates from globular
clusters or with the indirect age estimates, as provided by SNe Ia
and CMB measurements (Tonry 2002; Spergel et al. 2003).

On the other hand, it also well known that the evolution of the
age of the Universe with redshift ($dt_{\rm{U}}/dz$) differs from scenario to
scenario, which means that models that are old enough to explain  the total
expanding age at $z = 0$ may not be compatible with age estimates of
high-redshift objects. 
This in turn reinforces the idea that dating of objects at high-redshift
constitutes one of the most powerful methods for constraining the age of the
Universe at different stages of its evolution
(Dunlop et al. 1996; Kennicutt Jr. 1996; Spinrad et al. 1997), 
and the first epoch of galaxy/quasar formation
(Alcaniz \& Lima 2001; Alcaniz et al. 2003) as well as
for discriminating among different dark energy scenarios
(Krauss 1996; Alcaniz \& Lima 1999; Jimenez \& Loeb 2002; Jimenez et al. 2003).

In this concern, the recent discovery of an old quasar at a
redshift of $z = 3.91$ is of particular interest (Hasinger {\it et
al.} 2002; Komossa \& Hasinger 2002). Given its location at very
high redshift, the existence of APM 08279+5255, whose age was
firstly estimated to lie within the range of 2-3 Gyr (Hasinger
{\it et al.} 2002), it is of fundamental importance to study the
effect of any dark energy candidate and/or of the main
cosmological paramerers on the age of the Universe at the early
stages of the cosmological evolution (Komossa \& Hasinger 2002;
Alcaniz {\it{et al.}} 2003; Cunha \& Santos 2004). In principle,
provided that reliable and converging estimates on the inferred
age of APM 08279+5255 system are obtained, one may combine such
estimates with independent determinations of the matter density
parameter, $\Omega_m$, compare the predictions of different
scenarios and, possibly, explore new alternatives to better
describe the Universe. This is our primary objective in this
paper. Firstly, by using a chemodynamical model for the evolution
of spheroids proposed by Fria\c ca \& Terlevich (1998)
[hereafter FT98], we re-estimate the age for APM 0879+5255.
Further, by considering the current value of the matter density
parameter as given by WMAP team (Spergel et al. 2003), i.e.,
$\Omega_m = 0.27 \pm 0.04$ (1$\sigma$), and the latest
measurements of the Hubble parameter from the HST key project,
this new age estimate is used to test the viability of several
cosmological scenarios. It is shown that for these widely accepted
values of the matter density parameter ($\Omega_m = 0.27 \pm
0.04$) and of the Hubble parameter ($H_o = 72 \pm 8$
$\rm{km.s^{-1}.Mpc^{-1}}$) no dark energy scenario may explain the
estimated age for the APM 0879+5255 system.

\section{The age of APM 08279+5255 revisited}

\subsection{Chemical abundances constraints on the ages of high redshift objects}

It seems natural to consider the metallicity of a galaxy as an age indicator.
During its evolution, the galactic system becomes more metal-rich
in virtue of the metal injection by evolved stars, and, therefore,
a more metal-rich system would be expected to be older.
However, the star formation rate (SFR)
affects the relation between metallicity and age of a system,
because a higher SFR implies a faster rise of the metal
abundance. In this case, a relatively young object, that has been
formed at a very high SFR ($\sim 10$ Gyr$^{-1}$) could have high
metallicities. In particular, this explains why  even the highest
redshift quasars have strong metal lines in their spectrum.

Abundance ratios would be a better probe of galaxy age in chemical
evolution studies. Unlike absolute values, abundance ratios do not
strongly depend on particular model parameters and on details of
the star formation history, but mainly on the stellar
nucleosynthesis and the adopted initial mass function (IMF). 
In particular, the $\alpha$/Fe abundance ratio can be used
to place constraints on time scales for star formation and metal production.
While the $\alpha$ elements (O, Ne, Mg, Si, S)
are mainly produced in type II Supernovae (SNe II) in
short time scales ($\la 0.1$ Gyr), the Fe-peak elements are
produced in type Ia Supernovae (SNe Ia) in a longer time scale
($\sim 0.3-1$ Gyr). A slow (long time scale) star formation corresponds
to nearly solar values of $\alpha$/Fe while suprasolar values may
indicate shorter time scales for the star formation, which is
dominated by SNe II. Conversely, suprasolar values of the
Fe/$\alpha$ ratio indicate that the system is probably older than
1 Gyr.

Using an Fe/O=3 abundance ratio
(here the abundance ratio has been normalized to the solar value), 
derived from X-ray observations,
Komossa  and Hasinger (2002) estimated the age of the quasar APM 0879+5255
to lie within the interval 2-3 Gyr.
An age of 3 Gyr is inferred from the temporal
evolution of the Fe/O ratio in the giant elliptical model (M4a) of
Hamman \& Ferland (1993, hereafter HF93).
For the ``extreme model" M6a of HF93,
the Fe/O evolution would be faster, and Fe/O=3  is already reached after 2 Gyr.

It is worth to notice that the HF93 models are one-zone chemical
evolution models based on the classical monolithic scenario of
formation of elliptical galaxies. However, this scenario does not account for
internal variations in the SFR, which are required to explain the
metallicity (and colour) gradients of elliptical galaxies 
(Saglia et al. 2000; La Barbera et al. 2003),
and the internal variations of colours of spheroids
(Menanteau, Abrahams \& Ellis 2001;
Menanteau, Jimenez \& Matteucci 2001).
In addition, the
one-zone chemical evolution models that attempt to explain the
high metal content in high redshift QSOs tend to overproduce
metals (averaged over the entire galaxy) and predict an
excessively high luminosity for the parent galaxy of the QSO. For
example, the HF93 model M4a predicts luminosities of up to
$10^{15}$ L$_{\odot}$ during the formation of a present-day
$\sim$L$^*$ elliptical.

In what follows, we reevaluate the ages for APM 0879+5255 by
using a chemodynamical model for the evolution of spheroids (FT98).
We assume the scenario (which is the same adopted by HF93) in which luminous
quasars at high redshifts ($z \la 3$) are hosted in young
elliptical galaxies or massive spheroids (Dunlop et al. 2003).

\begin{figure}
\centerline{\psfig{figure=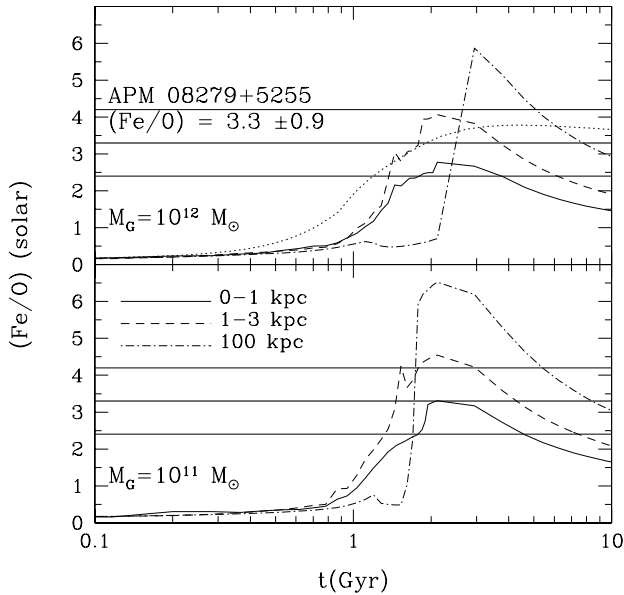,width=8.5cm}
\hskip 0.1in}
\caption{Top panel: evolution of the Fe/O ratio averaged over the
$r=0-1$ kpc and $r=1-3$ kpc spherical zones. 
The top panel
exhibits the predictions of the model with $M_G=10^{12}$ \msun, 
and the lower panel those for $M_G=10^{11}$
\msun. The horizontal lines denote the iron overbundance of Fe/O
of $3.3\pm 0.9$ found by Hasinger et al. 2002.
Also shown the Fe/O ratio in the halo of the galaxies,
at $r=100$ kpc.
In addition, the top panel exhibits the evolution of the Fe/O ratio
in the reference one-zone model of FT98 for a
$M_G=10^{12}$ \msun\ elliptical galaxy (dotted line).}
\end{figure}

\subsection{The chemodynamical model}

We investigate the joint chemical evolution of the quasar and its host galaxy
with the chemodynamical model of FT98, in which a single massive dark halo
hosts baryonic gas that will fall toward the centre of dark halo
and will subsequently form stars.
The code combines a multi-zone chemical evolution solver and a 1D
hydrodynamical code.
The system, assumed to be spherical, is
subdivided in several spherical zones and the
hydrodynamical evolution of its ISM is
calculated. The equations of chemical evolution for each
zone are then solved taking into account the gas flow,
and the evolution of the chemical abundances is obtained.
A total of $\approx$100 star generations are stored during 13 Gyr
for chemical evolution calculations.
We assume that at a given radius $r$ and the time $t$,
the specific SFR $\nu_{SF}$
follows a power-law function of gas density ($\rho$):
$\nu_{SF}(r,t) = \nu_0 (\rho/\rho_0)^{1/2}$,
where $\rho_0$ is the initial average gas density inside the core radius of the
dark halo ($r_h$), and $\nu$ is the normalization of the star formation law.

The stars formed follow a Salpeter IMF between 0.1 and 100 \msun.
Chemical evolution occurs as the stars formed out of the ISM
evolve and eject gas back into the ISM.
The stars are assumed to die either as supernovae (SNe) or as planetary 
nebulae, when instantaneous ejection of mass, metals and energy occurs.
The evolution of the abundances of He, C, N, O, Mg, Si, S, Ca and Fe
is calculated by solving the basic equations of chemical evolution.
We do not assume instantaneous recycling approximation for the chemical
enrichment, but instead we take into account
the delays for gas restoring from the stars
due to the main-sequence lifetimes.
Instantaneous mixing with the ISM is assumed for the stellar ejecta.
The models start with an entirely  gaseous protogalaxy
with primordial chemical abundances ($Y=0.24$, $Z=0$).
In this work, we use metallicity dependent yields
for SNe II, SNe Ia, and intermediate mass stars ($0.8 - 8$ \msun).
For more details of the nucleosynthesis prescriptions, see FT98
and Lanfranchi \& Fria\c ca (2003).

With the aid of this model, FT98 investigated
the relation between young elliptical galaxies and QSO activity.
It was used a sequence of galaxy models parameterized
according to the initial total baryonic mass $M_G=M_g+M_*$
(i.e. the stellar and the gas components).
The parameters of the model were set in order to reproduce
the present-day elliptical galaxy properties.
The relevant mass range for massive elliptical hosting luminous quasar is
$M_G=10^{11}-2\times10^{12}$ \msun.
Due to inflow and galatic wind episodes occuring during the galaxy evolution,
the present stellar mass of the galaxy is $\sim20$\% higher.

\subsection{Dating APM 0879+5255 with the chemodynamical model}

The chemodynamical model has been used to derive ages for
a variety of high redshift objects -- quasars (FT98),
elliptical galaxies in deep surveys (Jimenez et al. 1999),
Lyman Break Galaxies (Fria\c ca $\&$ Terlevich 1999),
Blue Core Spheroids (Fria\c ca $\&$ Terlevich 2001),
radiogalaxies/submm sources (Archibald et al. 2002),
and damped Lyman alpha systems (Lanfranchi \& Fria\c ca 2003) --
using several age indices: overall metallicity, level of QSO activity
IR-visible colours, colour gradients, IR-submm colours, abundance ratios.
Here we use the abundance ratio Fe/O as a clock of the
evolution of the galaxy.

From {\it XMM-Newton} observations of the high-redshift ($z=3.91$), lensed,
broad absorption line (BAL) QSO APM 0879+5255,
Hasinger et al. (2002) have derived an iron overbundance of Fe/O
of $3.3\pm 0.9$ (relative to solar abundances) for the BAL system.
The BAL system should be located near the centre of the galaxy,
where a supermassive black hole is driving the quasar activity. 
Figure 1 shows, at several radii, the evolution of the Fe/O ratio
(normalized to the solar abundances) for the models with
$M_G=10^{12}$ \msun\ and $M_G=10^{11}$ \msun.
This interval brackets most of the masses of galaxy hosts
of radio-loud and luminous QSOs (FT98, Archibald et al. 2002).
An age for the QSO of 2.11 Gyr is
set by the condition that Fe/O abundance ratio of the model reaches 3.3,
which corresponds to the best fit value given by Hasinger et al. (2002).
As we see from Figure 1, the age implied by Fe/O=3.3 
could be smaller in the outer regions of the galaxy, since in the very centre 
($r \la 1$ kpc), there is ongoing star formation for $\sim 2$ Gyr 
that gives rise to $\alpha$ element ejecta from SNe II, while in
the outer regions (e.g $r=1-3$ kpc), most of the star formation has ceased
before 1 Gyr
and there is metal ejection only from SNe Ia, leading to a
faster rise of the Fe/O ratio. However, the outer regions could
not host BALs, which are located only very close to the central
supermassive black hole.
Note that the model with $M_G=10^{12}$ \msun\ never reaches Fe/O=3.3.
However, the peak Fe/O abundance ratio (=2.77) also occurs at 2.1 Gyr.

The age predictions of one-zone models, as represented by the HF93 models,
are also in agreement with the results of the present, more elaborated model,
since Fe/O $\sim 3$ is reached at $\sim 2$ and $\sim 3$ Gyr by their models
M6a and M4a, respectively. Model M4 is the favorite QSO model of HF93,
whereas M6 is characterized as ``extreme", since its IMF
has a lower mass cutoff of 2.5 \msun.
In order to make a more detailed comparison of the chemodynamical
model to the one-zone model, 
Figure 1 also exhibits the predicted Fe/O evolution
of the reference one-zone model for elliptical galaxies of FT98.
This model represents an initially gaseous galaxy
with mass $M_G=10^{12}$ \msun, and no dark halo. 
It is assumed a linear star formation law, i.e. $\nu_{SF}=\nu_0=$ constant,
with $\nu_0=9$ Gyr$^{-1}$.
The model belongs to the class of SN-driven galactic wind models, in which
a galactic wind is established at a time $t_{gw}$.
At this time, the gas is swept from the galaxy and 
the star formation is turned off forever.
The condition for the onset of the galactic wind  is that
the thermal content of the SN remnants of the galaxy  equals
the gravitational binding energy of the galaxy ISM.
This model is similar to the model M4 of the HF93,
with the main differences being the lower $\nu_0=6.7$ Gyr$^{-1}$
and the flatter IMF (IMF power law slope $x=1.1$) of HF93.
In the FT98 one-zone reference model, the evolution of
the Fe/O is typical of the classic galactic wind models:
Fe/O is initially low, due to the predominant $\alpha$-element
enrichment by SNe II.
When the SN Ia appear, their Fe-rich ejecta boost the Fe/O ratio
which becomes solar at 0.7 Gyr. 
The Fe/O ratio keeps rising and, at $t_{gw}=0.81$ Gyr,
the galactic wind is established.
At later times, Fe/O reaches a plateau, with a slight decrease
as the model evolves, due to an increasing contribution of mass-loss
from intermediate mass stars.
In this model, Fe/O=3.3 is reached at 1.93 Gyr, somewhat before
the plateau phase of Fe/O evolution.

It is outstanding that both the chemodynamical and the one-zone models
predict that Fe/O $\sim 3$ is reached at $\sim 2$ Gyr,
even though the evolution of the Fe/O ratio is very different in the two models.
In order to address the issue of the 
coincidence of time scales for high iron enrichment in both models
and the evolution of the Fe/O ratio in the chemodynamical model,
we compare the central ($r<1$ kpc) iron enrichment
to the evolution of the SN Ia rate.
The first SN Ia appear at $3\times 10^7$ yr, but the SN Ia rate reaches
its peak only at 0.78 Gyr for model with $M_G=10^{11}$ \msun\ and
at 0.86 Gyr for that with $M_G=10^{12}$ \msun.
After the maximum, the decrease of the SN Ia rate is slow until $\sim 2$ Gyr,
when it is halved with respect to the peak value.
This behaviour is what is expected from a continuous star formation
lasting for $\sim 1$ Gyr.
In contrast, an instantaneous starburst
exhibits the maximum of SN Ia rate only $\sim 0.3$ Gyr after the burst.

As it appears, a continuous star formation history (for $\sim 1$ Gyr)
describes better the early evolution of the elliptical galaxy
than an instantaneous burst,
since in our model the main body of the galaxy is formed in $\sim 0.5$ Gyr,
but there is residual star formation in the central region
extending for the first $2-3$ Gyr of the galaxy evolution.
This is an important characteristic of the FT98 model, namely, that the
global star formation history of the elliptical galaxy is not coordinated
with the central star formation.

Not only the star formation, but also the gas flow 
is uncoordinated along the galaxy radius.
Some stages of the galaxy evolution are characterized by a complex flow pattern,
with inflow in some regions and outflow in other regions.
All models, however, exhibit during their late evolution a galactic wind
at the outer boundary and, during their early evolution, an inflow towards
the galaxy nucleus.
The central inflow maintains an extended star formation period
in the center and at the same time promotes the growth of the
QSO supermassive black hole and feeds its activity.
The length of the central inflow episode ($2-3$ Gyr) regulates the durarion
of the QSO activity and of the central star formation.

As a consequence of the extended central star formation period,
our elliptical galaxy model exhibits three star formation stages: 
stage (1) during which the bulk of the stellar population is formed
(half of the present day stellar population inside 10 kpc is formed in 0.40 and
0.30 Gyr for the galaxies with $10^{11}$ and $10^{12}$ \msun, respectively);
stage (2), which corresponds to an
extended (2-3 Gyr) period of star formation in the core maintained by the
inner cooling flow. This stage exhibits more modest star-formation
rates than stage (1) and explains the central blue colours of some spheroids at
$z=0.5-1$ (Fria\c ca \& Terlevich 2001)
and the absence of very red elliptical galaxies in deep optical and
near-infrared surveys (Jimenez et al 1999). 
Finally, during stage (3) after $\sim 3$ Gyr, the galaxy evolves passively
(i.e. with very low levels of star formation).

The relatively slow evolution of the SN rate (peak rate at $0.7-0.9$ Gyr)
is not inconsistent with short time scales for element enrichment
as required by the conspicuous metal lines in the spectra of high redshift QSOs.
In fact, in the chemodynamical model,
a high-metallicity core ($r<1$ kpc) is rapidly built-up. For the gas,
solar metallicities are reached in $10^8$ yr for oxygen, 
and $3\times 10^8$ yr for iron.
However, central Fe/O solar values are reached only at $\sim 1$ Gyr
(at 1.0 and 1.1 Gyr 
for the $M_G=10^{11}$ and $10^{12}$ \msun\ models, respectively).
The continuing star formation in the central regions of the galaxies
injects, via SNe II, $\alpha$-elements in the ISM,
which tend to lower the [Fe/O] ratio.
In the outer regions of the galaxy, where the star formation has ceased sooner,
the SN Ia are more effective in enriching the ISM in iron,
and the Fe/O rises faster and to higher values.
Note that the galactic wind, which reaches $r=100$ kpc at $1.1-1.3$ Gyr,
initially carries only the metal products of the early enrichment of the ISM,
predominantly $\alpha$ elements, but after $1.7-2.5$ Gyr, the iron
produced by SNe Ia arrives to the outer boundary of the galaxy and is
expelled into the intergalactic medium.

As noted above, not all models exhibit a central Fe/O as high as 3.3.
However, all the models reach Fe/O=2.5, and
it would be more useful for dating purposes, to focus on the rising
part of the Fe/O ratio evolution and establish a lower limit for the age of
APM 0879+5255 by considering when the Fe/O abundance ratio reaches 2.5.
This happens at 1.83 and 2.03 Gyr,
for the models with $10^{11}$ and $10^{12}$ \msun, respectively.
In addition, due to the later
decrease of the Fe/O ratio, there is only a temporal window
during which the QSO will be seen as very iron-rich (Fe/O=2.5 or higher):
1.8-4.3 Gyr, and 2.0-3.4 Gyr for the $M_G=10^{11}$ and $10^{12}$ \msun\ models,
respectively. In fact, this temporal window could be narrower,
because the later time would rather be given by the condition that
the QSO is being fed by the inner cooling flow, i.e. $t\la 3$ Gyr
(FT98, Archibald et al. 2002).
On the other hand, the lower limit for the age of the QSO
is more interesting for cosmological purposes.
As we will see in the next section, both our ``best estimate" of 2.1 Gyr 
for the age of APM 0879+5255, and the lower limits 1.8-2.0 Gyr
impose strict constraints on cosmological models.

\begin{figure*}
\centerline{\psfig{figure=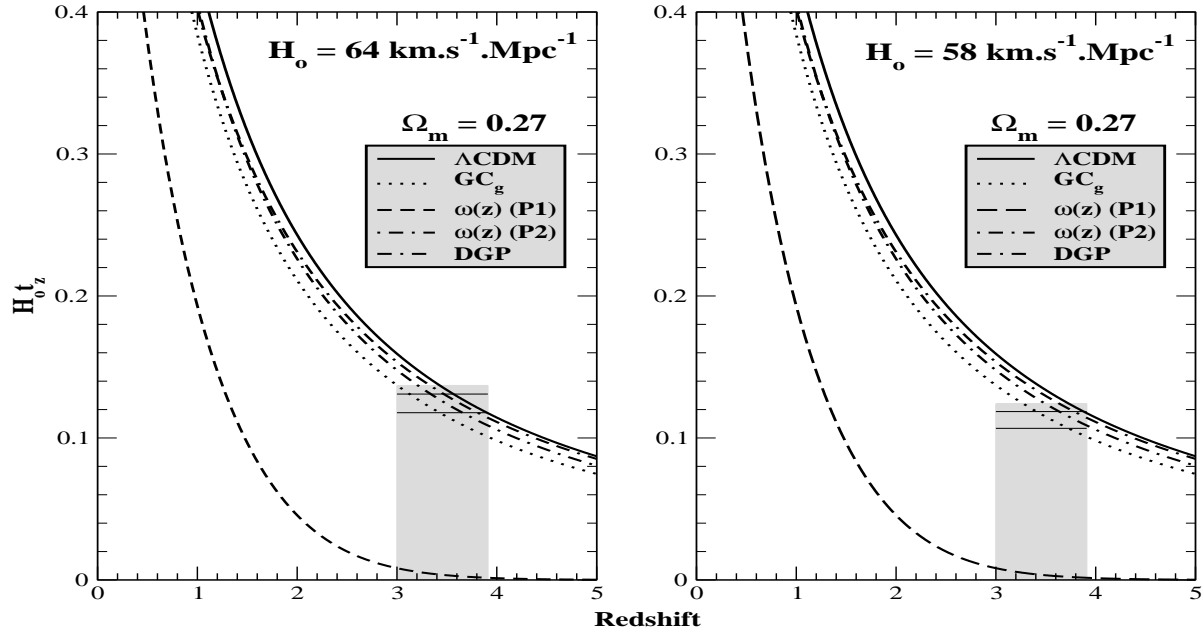,height=3.5truein,width=6.5truein,angle=-90}
\hskip 0.1in} \caption{Dimensionless age parameter as a function
of redshift for dark energy and brane world models. As explained
in the text, all curves crossing the shadowed area yield an age
parameter smaller than the value 2.1 Gyr required by the quasar
APM 08279+5255. 
The two thin horizontal lines in the
shadowed areas, correspond to the ages 1.8 and 2 Gyr, required for
the models with $10^{11}$ \msun\ and $10^{12}$ \msun\ to reach 
Fe/O=$2.5\times$ solar.
On the left panel, we see that none dark energy or
brane world scenario is compatible with the age estimate of 2.1 Gyr
for this object, even considering the 1$\sigma$ lower bound of the
Hubble parameter given by the HST key project. Note also that the same
happens for the central value, $H_o$ = 58 km.s$^{-1}$.Mpc$^{-1}$, as
recently advocated by Sandage and collaborators (right panel). 
}
\end{figure*}

\section{Testing cosmological scenarios with APM 0879+5255}

To what extent does this novel determination of the age of APM 0879+5255 provide new
constraints on dark energy scenarios? To answer this question, we first take for
granted that the age of the Universe at any redshift must be greater than or at least
equal to the age of its oldest objects. Such a condition naturally introduces the
following ratio (Alcaniz \& Lima 1999)
\begin{equation}
\frac{t_z}{t_g} = \frac{f(z; {\mathbf{P}})}{H_o t_g} \geq 1,
\end{equation}
where $t_z = H_{o}^{-1}f(z; {\mathbf{P}})$ is the predicted age of a particular
cosmological model, $t_g$ is the age of an arbitrary object, say, a quasar or a
galaxy at a given redshift $z$ and $f(z; {\mathbf{P}})$
is a dimensionless function of the cosmological parameters ${\mathbf{P}}$. For a given
object, the denominator of the above equation defines a dimensionless age parameter
$T_g = H_o t_g$, which involves two quantities measured or estimated from completely
independent methods. In particular, the 2.1-Gyr-old quasar at $z = 3.91$ yields $T_g =
2.1H_o$Gyr which, for the most
recent determinations of the Hubble parameter, $H_o = 72 \pm 8$
${\rm{km s^{-1} Mpc^{-1}}}$ (Freedman {\it et al.} 2001), takes
values on the interval $0.137 \leq T_g \leq 0.172$. In order to assure
the robustness of our analysis, we will adopt in our computations
the $1\sigma$ lower bound for the above mentioned value of the Hubble
parameter, i.e., $H_o = 64$ ${\rm{km s^{-1} Mpc^{-1}}}$, which implies $T_g \geq 0.131$.
Therefore, only models providing an age parameter larger than this value at $z =
3.91$ will be compatible with the existence of this object.

Figures 2 and 3 show the dimensionless age parameter $T_z =
H_ot_z$ as a function of the redshift for several dark energy
models, namely, the current concordance scenario, i.e., a flat
model with a cosmological constant ($\Lambda$CDM), the so-called
generalized Chaplygin gas [$\rm{GC_{g}}$] (Kamenshchik et al.
2001; Bili\'{c} et al. 2001; Bento et al. 2002; Dev et al. 2003),
two different parameterizations for the redshift-dependence of the
dark energy equation of state (EOS), i.e., $\omega(z) = \omega +
\omega_1z$ [P1] (Cooray \& Huterer 2000; Goliah et al. 2001) and
$\omega(z) = \omega + \omega_1(z/1+z)$ [P2] (Linder 2003;
Padmanabhan \& Choudhury 2003), and a 5-dimensional brane world
scenario, which we refer to it as DGP model (Dvali et al. 2000;
Deffayet et al. 2002; Alcaniz 2002). For the $\rm{GC_{g}}$ model
we assume $\alpha = 0.96$ and $A_s = 0.98$, the best-fit values
provided by the latest SNe Ia data (Alcaniz \& Lima 2004) while
for the $\omega(z)$ scenarios we fix $\omega = -1.31$ and
$\omega_1 = 1.48$, in agreement with the SNe Ia analysis of Riess
et al. (2004). The shadowed regions in the graphs are determined
from the minimal value of $T_g$, which means that any model
crossing the rectangles yields an age parameter smaller than the
minimal value required by the existence of of the quasar APM
08279+5255.

\begin{figure*}
\centerline{\psfig{figure=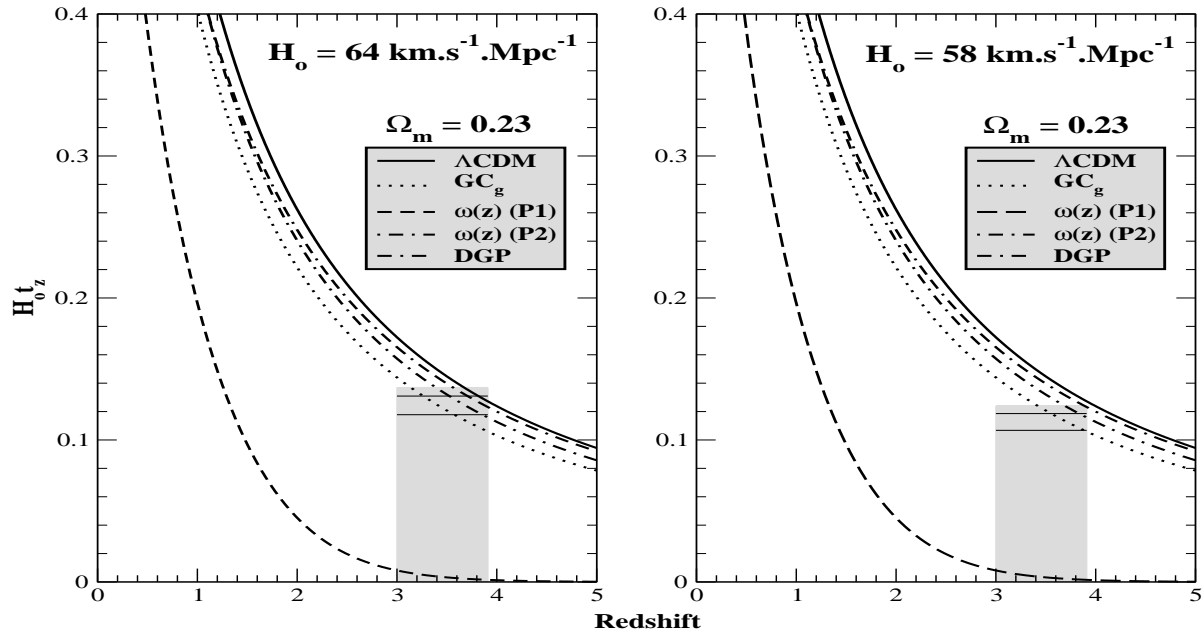,height=3.5truein,width=6.5truein,angle=-90}
\hskip 0.1in} \caption{The same as in Figure 2, but now for a
smaller value of the matter density parameter ($\Omega_m = 0.23)$.
We see that for the higher value of $H_o$ all the models cannot
accomodate the existence of the APM 08279+5255 system if $t_g=2.1$ Gyr (left
panel). However, for the central value advocated by Sandage and
collaborators, the ${\Lambda}$CDM and the DGP brane world scenario
are compatible (right panel). Note also that for any value of the
redshift the age predicted by the $\Lambda$CDM  is slightly
greater than the one for DGP brane world scenario.}
\end{figure*}

By adopting the central value of the matter density parameter as
given by WMAP team, i.e., $\Omega_m = 0.27$, we see from Figure 2a
that no dark energy or brane world scenario is compatible with the
age estimate of the APM 08279+5255 system for $H_o = 64 {\rm{km
s^{-1} Mpc^{-1}}}$, the $1\sigma$ lower bound for the current
accepted value of the Hubble parameter (Freedman {\it et al.}
2001). As shown in Figure 2b, such a situation persists even for
$H_o = 58 {\rm{km s^{-1} Mpc^{-1}}}$, the central value obtained
by Saha et al. (2001) and Sandage (2002). In particular, we note
that models parameterized like P1 for the redshift-dependence of
the dark energy EOS are in fully disagreement with any age
estimate at $z > 3.0$ being, therefore, ruled out as a possible
dark energy candidate. In Figure 3 a similar analysis is shown
having the matter density parameter fixed at $\Omega_m =
0.23$, which corresponds to 1$\sigma$ deviation below the best fit
value provided by WMAP (Spergel et al. 2003). Note that even in
this case, the unique way to make some scenarios compatible with
the estimated age for the APM 0879+5255 is admitting a value for
the Hubble parameter as low as $H_o = 58 {\rm{km s^{-1}
Mpc^{-1}}}$ (Sandage 2002). For this value of $H_o$, the age
parameter at $z = 3.91$ is $T_g \geq 0.124$ and both $\Lambda$CDM
and DGP models are compatible with the existence of the APM
0879+5255 system. By considering $H_o = 72 {\rm{km s^{-1}
Mpc^{-1}}}$, the central value of the HST key project (Freedman
{\it et al.} 2001) and no prior on the value of $\Omega_m$, we
find $\Omega_{\Lambda} \geq 0.85$ ($\Omega_m \leq 0.15$). Such
values are in disagreement with the results inferred from an
elementary combination of CMB measurements pointing to
$\Omega_{\rm{Total}} \simeq 1.0$ (de Bernardis et al. 2000;
Spergel et al. 2003) and clustering
estimates giving $\Omega_{\rm{m}} = 0.3 \pm 0.1$ (Calberg {\it et
al.} 1996; Dekel {\it et al.} 1997) or from a rigorous statistical
analysis involving many astrophysical constraints (see, for
example, Peebles \& Ratra (2002), Padmanabhan (2002) and Lima (2004)
for recent reviews).  
For the other three scenarios considered in this paper (excluding P1),
we obtain: 
$\Omega_m < 0.1$ (P2), $\Omega_m < 0.16$ (Cg) [for values of $A_s = 1.0$,
$\alpha = 0$, which correspond to the maximum age] and $\Omega_m < 0.15$ (DGP).

\section{Summary and Conclusions}

As widely known, the determination of the total age of the universe (the expanding
time from the big-bang to $z=0$) has been since the early thirties one of the major
questions as well as a real source of progress for
cosmology. Similarly, the age estimates of old high redshift
objects may play a prominent role to discriminate among the
existing dark energy or brane world models by constraining the
basic cosmological parameters. In reality, the so-called high
redshift ``age crisis" seems to be even more restrictive than the
total age, and is now becoming an important complement to other
independent cosmological tests (Alcaniz \& Lima 1999; Jimenez \& Loeb 2002).

In this paper we have reestimated  the age of the APM 08279+5255 quasar, 
and constrained some basic cosmological
parameters of a large class of cosmological scenarios. 
In the first place, 
using a chemodinamical model for the evolution of spheroids, 
we reevaluate its current estimated age, as
given by Hasinger {\it et al.} (2002). 
An age of 2.1 Gyr is set by the condition that 
Fe/O abundance ratio (normalized to solar values) of the model reaches 3.3,
which is the best fit value given in Hasinger {\it et al.} (2002).
In the detailed chemodynamical modelling, although the central
region of the galaxy housing the QSO reaches a solar iron abundance
in $\sim 0.3$ Gyr, the Fe/O abundance ratio reaches the solar value
at $\sim 1$ Gyr, and Fe/O=2.5 only at $\sim 2$ Gyr.
Therefore, a highly suprasolar value of the Fe/O abundance ratio for a QSO
is a strong evidence that the QSO is old,
which constrains severely cosmological scenarios. 

As we have seen (Figures 2 and 3), assuming $t_g=2.1$ Gyr for APM 08279+5255,
for the current accepted values of
$\Omega_{\rm{m}}$, the main class of world models (dark energy or
brane inspired universes) cannot accomodate the existence of this
object. $\Lambda$CDM and DGP brane world models pass the test only
if one considers the low value of $H_o$ obtained by Sandage and
collaborators (Figure 3). 
Even less stringent age limits, derived from a Fe/O ratio of 2.5, 
which is allowed by the X-ray data, 
are consistent with only a few  cosmological scenarios,
namely, those favoring an older high-redshift Universe.
The models with $10^{11}$ \msun\ and $10^{12}$ \msun\ reach 
Fe/O=$2.5\times$ solar at 1.8 and 2.0 Gyr, respectively.
In this case (see Figures 2 and 3), also the 
$\Lambda$CDM and DGP brane world models are favored
by the age estimates.
In addition, $H_o = 64 {\rm{kms^{-1} Mpc^{-1}}}$
could be acceptable if  $\Omega_m = 0.23$.

In general grounds, two aspects should be
emphasized. First, by using an independent chemodynamical
avolution code, we have confirmed the earlier age estimates of
this object set by Hasinger et al (2002). Second, we have shown
the importance of the dating of high redshift objects as an
independent test. This is more evident, for instance, in the case
of $\omega(z)$ (P1) scenario which predicts a total expanding age
compatible with the current dating of globular clusters (Asantha
\& Huterer 1999), but which present a very small age at moderate
and high $z$ (see the dashed line in Figures 2 and 3). Our results,
therefore, rule out the model as a realistic  description of our
Universe.

The present work highlights the cosmological interest
on the observational search for quasars, galaxies
and other old collapsed objects  at high redshifts.
In particular, we emphasize that chemical modelling of high redshift quasar
evolution can be a valuable tool for dating the high redshift Universe,
providing much needed constraints on cosmological models.

\section*{Acknowledgments}
This work is supported by the Conselho Nacional de Desenvolvimento
Cient\'{\i}fico e Tecnol\'{o}gico (CNPq - Brazil) and CNPq
(62.0053/01-1-PADCT III/Milenio).

\label{lastpage}

\end{document}